\documentclass[]{aa}  

\usepackage{graphicx}
\usepackage[varg]{txfonts}
\usepackage{color}
\usepackage{ulem}

\begin{document}

   \title{On the Number of Stars in the Sun's Birth Cluster}

   \author{Sota Arakawa
          \inst{1,2}
          \and
          Eiichiro Kokubo
          \inst{2}
          }

   \institute{Japan Agency for Marine-Earth Science and Technology, \\
             3173-25, Showa-machi, Kanazawa-ku, Yokohama, Kanagawa 236-0001, Japan\\
             \email{arakawas@jamstec.go.jp}
         \and
             National Astoronomical Observatory of Japan, 2-21-1 Osawa, Mitaka, Tokyo 181-8588, Japan
             }

   \date{Received date / Accepted date}

% \abstract{}{}{}{}{} 
% 5 {} token are mandatory
 
  \abstract{
The Sun is thought to be formed within a star cluster.
The coexistence of $^{26}{\rm Al}$-rich and $^{26}{\rm Al}$-poor calcium--aluminum-rich inclusions indicates that a direct injection of  $^{26}{\rm Al}$-rich materials from a nearby core-collapse supernova should occur in the first $10^5$ years of the solar system.
Therefore, at least one core-collapse supernova should occur within the duration of star formation in the Sun's birth cluster.
Here we revisit the number of stars in the Sun's birth cluster from the point of view of the probability for acquiring at least one core-collapse supernova within the finite duration of star formation in the birth cluster.
We find that the number of stars in the birth cluster can be significantly larger than that previously considered, depending on the duration of star formation.
}


   \maketitle
%
%-------------------------------------------------------------------

\section{Introduction}

The majority of stars form within star clusters \citep[e.g.,][]{2003ARA&A..41...57L}, and our solar system is also thought to be formed within a cluster from many perspectives \citep[e.g.,][]{2010ARA&A..48...47A}.
For example, some of trans-Neptunian objects including (90377) Sedna have large heliocentric eccentricity and perihelion distance \citep[e.g.,][]{2004ApJ...617..645B}, and the plausible scenario for their formation is a close encounter with a passing star in the Sun's birth cluster \citep[e.g.,][]{2004AJ....128.2564M, 2006Icar..184...59B}.
Parts of presolar grains, whose isotopic compositions are significantly anomalous by comparison with the average solar composition, originate from core-collapse supernovae (CCSNe; e.g., \cite{2016ARA&A..54...53N}), and the chromium isotopic heterogeneity found in various chondritic meteorites might be the evidence of the injection of dust grains from a CCSN \citep[e.g.,][]{2014M&PS...49..772S, 2021ApJ...908...64F}.

Injection of dust grains into the solar system from a nearby CCSN is also a classical scenario for the origin of short-lived radionuclides (SLRs) found in meteorites \citep[e.g.,][]{2008ApJ...688.1382T, 2010ApJ...711..597O}.
SLRs are radioactive nuclides that were present in the early solar system but now extinguished, and their existence in the early solar system is recorded in meteorites as excesses of the daughter isotopes.
\citet{2009GeCoA..73.4922H} reviewed the stellar nucleosynthetic processes in the context of stellar evolution, and they concluded that massive stars in the range of $20$--$60 m_{\sun}$ ($m_{\sun}$ is the solar mass) would be plausible sources of SLRs when we assumed the direct injection from CCSNe.
The range of the progenitor mass was discussed based on the amount of production for several SLRs including $^{26}{\rm Al}$ and $^{60}{\rm Fe}$.

The oldest dust particles condensed in the solar system are calcium--aluminum-rich inclusions (CAIs) 4.567 billion years ago based on their lead isotopic ages \citep[e.g.,][]{2002Sci...297.1678A}.
Although the majority of primitive (i.e., unmelted) CAIs are uniformly enriched in $^{26}{\rm Al}$ \citep[e.g.,][]{2012E&PSL.331...43M}, some unusual CAIs show the order-of-magnitude lower initial abundance of $^{26}{\rm Al}$.
Platy hibonite crystals (PLACs) are classified into those unusual CAIs, and they are regarded as the oldest CAIs based on their high condensation temperature and large nucleosynthetic anomalies \citep[e.g.,][]{1998ApJ...509L.137S, 2016GeCoA.189...70K}, although their origin is still under debate \citep[e.g.,][]{2020E&PSL.53516088L, 2022arXiv220311169D}.
CAIs with fractionation and unidentified nuclear effects (FUN CAIs) have also low and varied initial abundance of $^{26}{\rm Al}$, and they are thought to be formed prior to normal CAIs \citep[e.g.,][]{2017GeCoA.201....6P}.
The coexistence of $^{26}{\rm Al}$-rich and $^{26}{\rm Al}$-poor CAIs would be the evidence of the direct injection of $^{26}{\rm Al}$-enriched dust grains into the solar nebula (or the solar molecular cloud core) in the epoch of CAI formation \citep[e.g.,][]{1998ApJ...509L.137S, 2013PNAS..110.8819H}.
As the duration of CAI formation (i.e., the formation age spread of CAIs) is a few $10^{5}$ years or less \citep[e.g.,][]{2006ApJ...646L.159T, 2012Sci...338..651C, 2020GeCoA.279....1K}, the direct injection event would have occurred in the first $10^{5}$ years of the solar system formation.
We note that the initial abundance of $^{26}{\rm Al}$ recorded in CAIs is not altered even if additional injection of $^{26}{\rm Al}$-enriched materials into the solar nebula occurred after the condensation of CAIs.
This timescale is one order of magnitude shorter than the duration of star formation in star clusters \citep[e.g.,][]{2019Natur.569..519K, 2021PASJ...73.1074F}, and we can expect that at least one CCSN would occur in the Sun's birth cluster during its star formation period.

The condition for acquiring at least one CCSN in the birth cluster has been studied in previous studies \citep[e.g.,][]{2010ARA&A..48...47A, 2019A&A...622A..69P}.
They evaluated the minimum number of stars in the birth cluster based on the probability for existing at least one massive star whose mass is large enough to trigger a CCSN with ejection of SLR-rich dust grains.
For example, \citet{2010ARA&A..48...47A} demonstrated that the probability for hosting at least one massive star whose mass is larger than $25 m_{\sun}$ reaches 50\% when the number of stars in the cluster is approximately 800.
We note, however, that these estimates did not consider the timing of CCSNe in the cluster, and the existence of massive stars in the cluster does not support the occurrence of CCSN events during its star formation period.
This is because there is a finite time lag for the birth of a massive star and its explosion.
As both the time lag (i.e., the lifetime of a massive star) and the duration of star formation in a cluster are several to tens of million years \citep[e.g.,][]{1992A&AS...96..269S}, the lifetime of a massive star could have a large impact on the evaluation of the number of stars in the birth cluster.

\begin{figure*}
\begin{center}
\includegraphics[width=0.8\textwidth]{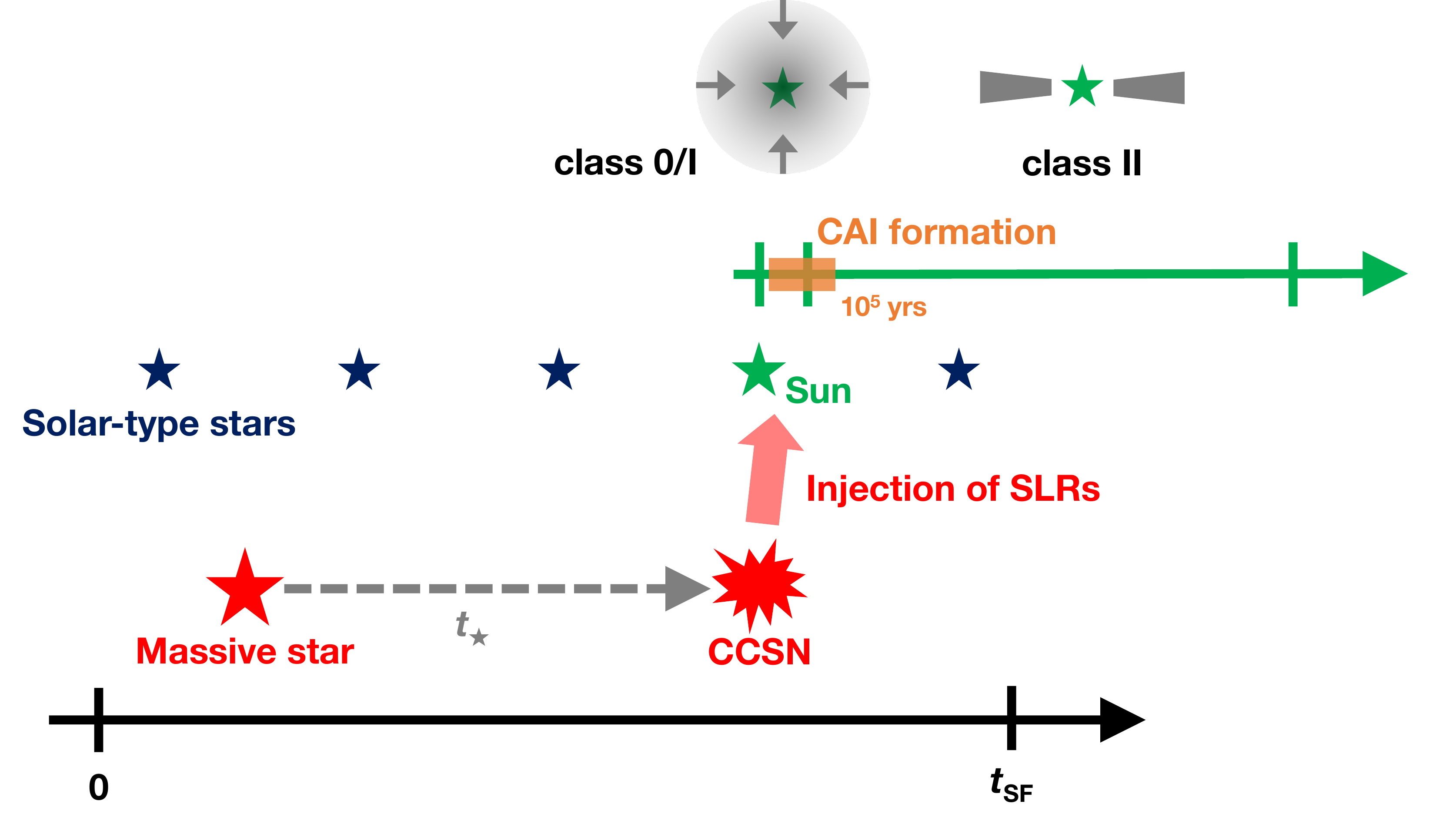}
\end{center}
\caption{
Schematic of the direct injection of SLRs into the early solar system within the birth cluster.
Massive stars those were born in the cluster would trigger CCSNe when they finished their lifetime, $t_{\star}$.
To explain the coexistence of $^{26}{\rm Al}$-rich and $^{26}{\rm Al}$-poor CAIs in the early solar system, a direct injection of SLR-rich dust grains from a CCSN should occur during CAI formation in the solar system.
The necessary condition to form the solar system is that at least one CCSN occurs in the birth cluster within the duration of star formation, $t_{\rm SF}$.
}
\label{fig.1}
\end{figure*}

Figure \ref{fig.1} shows the schematic of the direct injection of SLRs into the early solar system within the birth cluster.
Stars continuously formed within the birth cluster, and the duration of star formation, $t_{\rm SF}$, would be several million years.
Massive stars those were born in the cluster would trigger CCSNe when they finished their lifetime, $t_{\star}$.
A large number of solar-type stars also formed in the cluster.
To explain the coexistence of $^{26}{\rm Al}$-rich and $^{26}{\rm Al}$-poor CAIs in the early solar system, a direct injection of SLR-rich dust grains from a CCSN should occur during CAI formation in the solar system.
As the duration of CAI formation is a $10^{5}$ years and negligibly shorter than $t_{\rm SF}$, the necessary condition to form the solar system is that at least one CCSN occurs in the birth cluster within $t_{\rm SF}$.

In this paper, we revisit the number of stars in the Sun's birth cluster from the point of view of direct injection of SLRs from a CCSN to the early solar system.
We calculated the probability for acquiring at least one CCSN within the finite duration of star formation in the birth cluster.
We found that the estimated number of stars in the birth cluster is significantly larger than that previously considered, especially for the cases that the duration of star formation is far less than 10 million years; e.g., for $t_{\rm SF} = 5~{\rm Myr}$, the probability is 50\% when the number of stars in the cluster is approximately $2 \times 10^{4}$, which is 25 times larger than the estimated number by \citet{2010ARA&A..48...47A}.

\section{Model}

In this section, we briefly describe basic equations used in this study.
We calculate the expected number of CCSN events during the star formation period of the birth cluster using probabilistic approach.
Our model is similar to that considered in \citet{2010ARA&A..48...47A}.
The novel point of this study is that we consider the impacts of finite duration of star formation and lifetime of massive stars on the expected number of CCSN events.

\subsection{Initial mass function}

It is known that the maximum stellar mass in the birth cluster, $m_{\rm max}$, depends on the total cluster mass, $M_{\rm cl}$ \citep[e.g.,][]{2006MNRAS.365.1333W}.
\citet{2019A&A...622A..69P} provides an approximated relation between $m_{\rm max}$ and $M_{\rm cl}$ as follows:
\begin{eqnarray}
{\log_{10} \frac{m_{\rm max}}{m_{\sun}}} & = & - 0.76 + 1.06{\log_{10} \frac{M_{\rm cl}}{m_{\sun}}} \nonumber \\
                                           &   & - 0.09{\left( {\log_{10} \frac{M_{\rm cl}}{m_{\sun}}} \right)}^{2},
\label{eq.m_max}
\end{eqnarray}
where $m_{\sun}$ is the solar mass.
The minimum stellar mass, $m_{\rm min}$, is set to be $m_{\rm min} = 0.07 m_{\sun}$.

The initial mass function of stars in the birth cluster, ${\xi ( m_{\star} )}$, is given by a broken power-law mass function \citep[e.g.,][]{2002Sci...295...82K, 2021arXiv211210788K}, where $m_{\star}$ is the stellar mass.
For low-mass stars of $m_{\rm min} < m_{\star} < m_{\rm tra}$, the initial mass function is given by
\begin{equation}
{\xi ( m_{\star} )} = k {\left( \frac{m_{\star}}{m_{\rm min}} \right)}^{- 1.3},
\end{equation}
where $m_{\rm tra} = 0.5 m_{\sun}$ is the transition mass for the broken power-law and $k$ is the constant.
For high-mass stars of $m_{\rm tra} \leq m_{\star} < m_{\rm max}$, the initial mass function is given by
\begin{equation}
{\xi ( m_{\star} )} = k {\left( \frac{m_{\rm tra}}{m_{\rm min}} \right)}^{- 1.3} {\left( \frac{m_{\star}}{m_{\rm tra}} \right)}^{- 2.3}.
\end{equation}
The initial mass function satisfies the following equation:
\begin{equation}
\int_{m_{\rm min}}^{m_{\rm max}} {\rm d}m_{\star}~m_{\star} {\xi ( m_{\star} )} = M_{\rm cl}.
\end{equation}
The expected total number of stars in the birth cluster, $N_{\rm cl}$, is
\begin{equation}
\int_{m_{\rm min}}^{m_{\rm max}} {\rm d}m_{\star}~{\xi ( m_{\star} )} = N_{\rm cl}.
\end{equation}

\begin{figure}
\begin{center}
\includegraphics[width=0.48\textwidth]{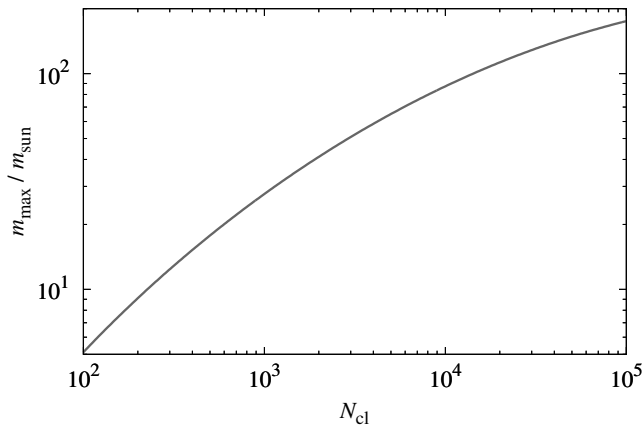}
\end{center}
\caption{
The relation between the maximum stellar mass in the birth cluster, $m_{\rm max}$, and the total number of stars, $N_{\rm cl}$.
}
\label{fig.M}
\end{figure}

Figure \ref{fig.M} shows the relation between $m_{\rm max}$ and $N_{\rm cl}$.
As massive stars of $20 m_{\sun} < m_{\star} < 60 m_{\sun}$ can trigger CCSNe that might provide SLRs to the early solar system, $N_{\rm cl}$ for a cluster that hosts a progenitor of a plausible CCSN is $N_{\rm cl} \gtrsim 600$.

\subsection{Time distribution of star formation}

Each star in the birth cluster forms at different time within the finite duration.
The formation rate of stars whose mass is $m_{\star}$ and the birth time is $t$, ${n_{\rm SF} ( m_{\star}, t )}$, is given by
\begin{equation}
{n_{\rm SF} ( m_{\star}, t )} = {\xi ( m_{\star} )} {f ( m_{\star}, t )},
\end{equation}
where ${f ( m_{\star}, t )}$ is the normalized time distribution of star formation, which is defined as
\begin{equation}
\int_{0}^{t_{\rm SF}} {\rm d}{t}~{f ( m_{\star}, t )} = 1,
\label{eq.f}
\end{equation}
and $t_{\rm SF}$ is the duration of star formation in the birth cluster.
In this study, we assume that ${f ( m_{\star}, t )}$ is given as the following mass- and time-independent equation for simplicity:
\begin{equation}
{f ( m_{\star}, t )} = \frac{1}{t_{\rm SF}}.
\label{eq.fb}
\end{equation}

We note that ${f ( m_{\star}, t )}$ must depend on both $m_{\star}$ and $t$ in reality.
Direct numerical simulations of the star cluster formation in giant molecular clouds \citep[e.g.,][]{2020MNRAS.497.3830F, 2021MNRAS.506.5512F} are necessary to discuss the detail of ${f ( m_{\star}, t )}$.
\citet{2021MNRAS.506.5512F} calculated the temporal evolution of the total stellar mass in their numerical simulations of star cluster formation.
The total stellar mass increased with time and saturated at several times of free-fall timescale of star-forming molecular clouds.
From their Figure 7, ${f ( m_{\star}, t )}$ takes the maximum at around the free-fall timescale, although the timing must depend on the physicochemical properties of star-forming molecular clouds including the initial density profile and the metallicity in reality (see Section \ref{sec.duration}).

\subsection{Lifetime of stars}

\begin{figure}
\begin{center}
\includegraphics[width=0.48\textwidth]{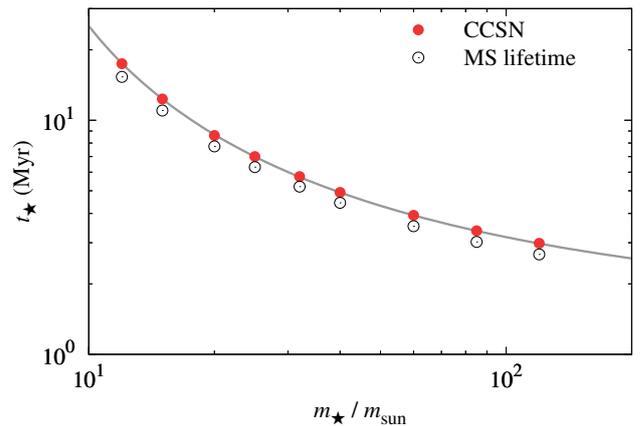}
\end{center}
\caption{
The relation between the lifetime of stars (i.e., the time lag from the birth of a star to trigger CCSN), $t_{\star}$, and the stellar mass, $m_{\star}$.
We also show the main-sequence lifetime (i.e., the duration of H-burning) as reference.
Circle plots are data taken from \citet{2012A&A...537A.146E}, and the solid curve is the empirical fitting.
}
\label{fig.LT}
\end{figure}

The lifetime of stars depends on the stellar mass \citep[e.g.,][]{1992A&AS...96..269S}.
Assuming that the lifetime of the star, $t_{\star}$, is given by the sum of durations of H-burning and He-burning, we obtained an empirical relation between $t_{\star}$ and $m_{\star}$ for non-rotating massive stars by using the numerical results shown in \citet{2012A&A...537A.146E};
\begin{equation}
\frac{t_{\star}}{1~{\rm Myr}} = 3.25 {\left( {\log_{10} \frac{ m_{\star}}{4.69 m_{\sun}}} \right)}^{-1.80} + 1.22.
\end{equation}
Figure \ref{fig.LT} shows $t_{\star}$ as a function of $m_{\star}$, and we found that $t_{\star} > 4~{\rm Myr}$ for massive stars of $20 m_{\sun} < m_{\star} < 60 m_{\sun}$, i.e., the progenitors of plausible CCSNe \citep[e.g.,][]{2009GeCoA..73.4922H}.
It is important to note that no CCSNe that provide SLRs to the early solar system would be triggered within $t_{\rm SF}$ if the lifetime of the most massive star in the cluster, $t_{\star, {\rm max}}$, is longer than $t_{\rm SF}$.

\section{Results}

When we assume that the situation that a CCSN occurred at the timing of the birth of the Solar System, we need (at least) one CCSN event within the duration of star formation in the birth cluster.
In this section, we discuss the probability for acquiring at least one CCSN within the duration of star formation.

\subsection{Probability for acquiring at least one CCSN within the duration of star formation}

The event rate of CCSNe whose progenitor mass is $m_{\star}$ and the explosion time is $t$, ${n_{\rm SN} ( m_{\star}, t )}$, is given by
\begin{equation}
{n_{\rm SN} ( m_{\star}, t )} = {n_{\rm SF} ( m_{\star}, t - t_{\star} )}.
\label{eq.nSN}
\end{equation}
Then the total event rate of CCSNe at $t$, ${p_{\rm SN} ( t )}$, is
\begin{equation}
{p_{\rm SN} ( t )} = \int_{20 m_{\sun}}^{60 m_{\sun}} {\rm d}m_{\star}~{n_{\rm SN} ( m_{\star}, t )}.
\end{equation}

\begin{figure}
\begin{center}
\includegraphics[width=0.48\textwidth]{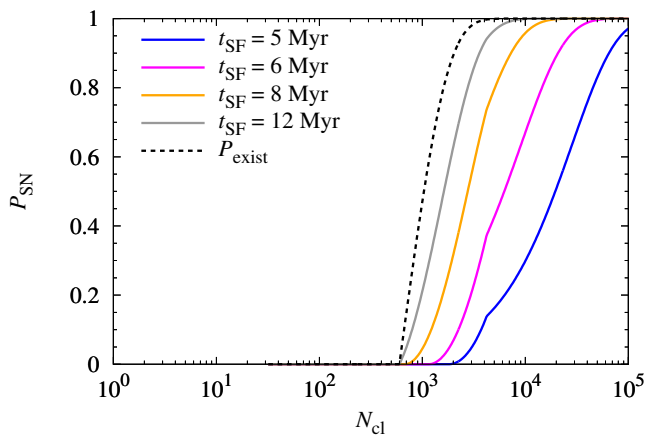}
\end{center}
\caption{
Probability for acquiring at least one core-collapse supernova with mass in the range of $20$--$60 m_{\sun}$ within the duration of star formation, $P_{\rm SN}$.
Here we assume that $m_{\rm max}$ depends on $N_{\rm cl}$ and it is given by Equation (\ref{eq.m_max}).
}
\label{fig.P}
\end{figure}

The cumulative number of CCSNe within the duration of star formation in the birth cluster, $Z_{\rm SN}$, is given as follows:
\begin{equation}
Z_{\rm SN} = \int_{0}^{t_{\rm SF}} {\rm d}{t}~{p_{\rm SN} ( t )}.
\end{equation}
Then the probability for acquiring at least one CCSN, $P_{\rm SN}$, would be given by (see Appendix \ref{appA})
\begin{equation}
P_{\rm SN} = 1 - \exp{\left( - Z_{\rm SN} \right)}.
\end{equation}
Figure \ref{fig.P} shows $P_{\rm SN}$ and its dependence on $N_{\rm cl}$ and $t_{\rm SF}$.
The minimum number of $N_{\rm cl}$ for $P_{\rm SN} > 0$ depends on $t_{\rm SF}$; this is because $t_{\star, {\rm max}}$ depends on $N_{\rm cl}$, and $P_{\rm SN} = 0$ when $t_{\rm SF} \leq t_{\star, {\rm max}}$.

The birth cluster of the Sun needs to satisfy the condition of $P_{\rm SN} > 0$ and probably meets the condition $P_{\rm SN} \gtrsim 0.5$ \citep[e.g.,][]{2010ARA&A..48...47A}.
We note that the threshold value of $P_{\rm SN} = 0.5$ was arbitrarily chosen in \citet{2010ARA&A..48...47A}, although 50\% would be a fair threshold from the perspective of whether at least one CCSN event is likely to occur within $t_{\rm SF}$ in a star cluster or not.
If $P_{\rm SN} \ll 1$ for a star cluster, no solar-type stars would have experienced a direct injection of SLR-rich materials in the star cluster.
We acknowledge, however, that one of solar-type stars could have experienced a direct injection even if $P_{\rm SN}$ for each star cluster is $P_{\rm SN} \ll 1$ when we consider dozens of star clusters.

Our results indicate that the total number of stars in the birth cluster would be $N_{\rm cl} \gtrsim 2 \times 10^{4}$ if $t_{\rm SF} = 5~{\rm Myr}$, although it depends on the shape of ${f ( m_{\star}, t )}$.
When the duration of star formation is $t_{\rm SF} = 12~{\rm Myr}$, the required condition for $N_{\rm cl}$ is mitigated: $N_{\rm cl} \gtrsim 2 \times 10^{3}$.

Considering the relation between timings of star formation and explosion (see Equation (\ref{eq.nSN})), $P_{\rm SN}$ depends on the shape of ${f ( m_{\star}, t )}$.
If ${f ( m_{\star}, t )}$ is low at $t \sim 0$ and increases with time for massive stars of $20$--$60 m_{\sun}$, $P_{\rm SN}$ is lower than that for time-independent ${f ( m_{\star}, t )}$.
Recent astronomical observations of nearby star-forming regions \citep[e.g.,][]{2020A&A...642A..87K, 2022A&A...658A.114K} revealed that low-mass stars were born in filamentary structures while massive stars were born in hubs formed via merging of multiple filaments.
We speculate that ${f ( m_{\star}, t )}$ is low at $t \sim 0$ for massive stars if star clusters formed from hub--filament systems, although it should be tested in future numerical simulations and astronomical observations.

\subsection{Impacts of the initial mass function on $P_{\rm SN}$}

In the previous section, we considered that $m_{\rm max}$ depends on $N_{\rm cl}$ as shown in Figure \ref{fig.M}.
However, some of previous studies did not consider this dependence but they assumed that $m_{\rm max}$ as a constant.
In \citet{2010ARA&A..48...47A}, $m_{\rm max}$ is set to $100 m_{\sun}$.
Here we show how the assumption of $m_{\rm max}$ affects the probability for acquiring at least one CCSN within the duration of star formation.

\begin{figure}
\begin{center}
\includegraphics[width=0.48\textwidth]{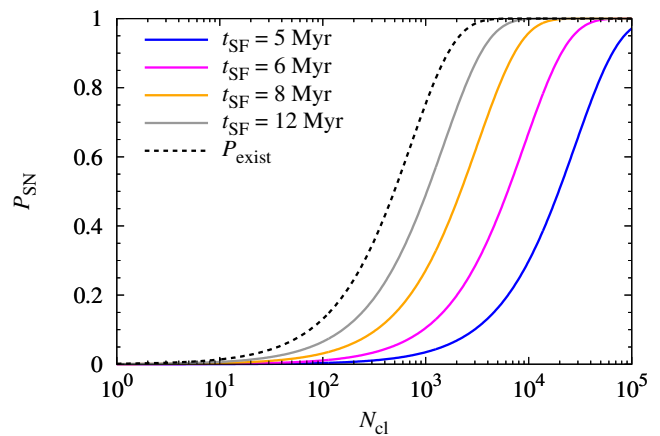}
\end{center}
\caption{
Probability for acquiring at least one core-collapse supernova with mass in the range of $20$--$60 m_{\sun}$ within the duration of star formation, $P_{\rm SN}$.
Here we assume that $m_{\rm max} = 100 m_{\sun}$ and it does not depend on $N_{\rm cl}$.
}
\label{fig.P_c}
\end{figure}

Figure \ref{fig.P_c} shows $P_{\rm SN}$ and its dependence on $N_{\rm cl}$ and $t_{\rm SF}$.
In contrast to Figure \ref{fig.P}, $P_{\rm SN}$ is nonzero for any $N_{\rm cl}$ when $t_{\rm SF}$ is longer than the lifetime of stars whose mass is $60 m_{\sun}$ ($\simeq 4~{\rm Myr}$).
We found that $P_{\rm SN}$ barely changes with the assumption of $m_{\rm max}$ for the cases of $N_{\rm cl} \gtrsim 4 \times 10^{3}$.
This $N_{\rm cl}$ corresponds to the condition for $m_{\rm max} = 60 m_{\sun}$.
As both mass and number of stars in the cluster is dominated by smaller stars, the number of massive stars with mass in the range of $20$--$60 m_{\sun}$ barely depends on the assumption of $m_{\rm max}$ when $m_{\rm max} > 60 m_{\sun}$.

We also calculate the probability for existing at least one massive star whose mass is large enough to trigger a plausible CCSN.
The expected number of progenitors formed in the birth cluster, $Z_{\rm exist}$, is given by
\begin{equation}
Z_{\rm exist} = \int_{20 m_{\sun}}^{60 m_{\sun}} {\rm d}m_{\star}~{\xi ( m_{\star} )}.
\end{equation}
Then the probability for existing at least one progenitor, $P_{\rm exist}$, would be given by
\begin{equation}
P_{\rm exist} = 1 - \exp{\left( - Z_{\rm exist} \right)}.
\end{equation}
We note that $P_{\rm SN} \to P_{\rm exist}$ for $t_{\rm SF} \to \infty$ by definition.

The classical estimation by \citet{2010ARA&A..48...47A} corresponds to the dashed line of Figure \ref{fig.P_c}, whereas our results those take into consideration the finite duration of star formation are the solid lines in  Figure \ref{fig.P}.
We found that $P_{\rm exist} > 0.5$ is achieved when $N_{\rm cl} > 500$ and $m_{\rm max}$ is independent of $N_{\rm cl}$.
In contrast, when we consider both finite $t_{\rm SF}$ and variable $m_{\rm max}$, $P_{\rm SN} > 0.5$ is achieved only when $N_{\rm cl} \gg 1000$.
For example, $N_{\rm cl} \gtrsim 2 \times 10^{4}$ if $t_{\rm SF} = 5~{\rm Myr}$ and this is 40 times larger than the value obtained from a classical way.
In particular, the impact of finite $t_{\rm SF}$ on the estimate of $N_{\rm cl}$ is significant as shown in Figures \ref{fig.P} and \ref{fig.P_c}.

\subsection{Plausible parameters of the Sun's birth cluster in $N_{\rm cl}$--$t_{\rm SF}$ plane}

\begin{figure}
\begin{center}
\includegraphics[width=0.48\textwidth]{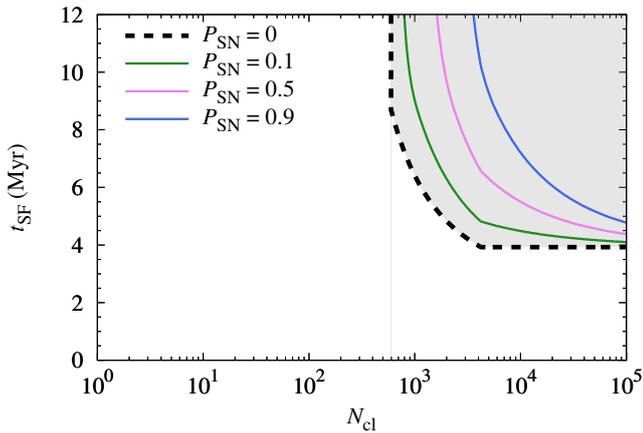}
\end{center}
\caption{
Contour chart of $P_{\rm SN}$ in $N_{\rm cl}$--$t_{\rm SF}$ plane.
We note that $P_{\rm SN} = 0$ outside the shaded area.
}
\label{fig.map}
\end{figure}

Figure \ref{fig.map} is the contour chart of $P_{\rm SN}$ in $N_{\rm cl}$--$t_{\rm SF}$ plane.
We found that $P_{\rm SN} = 0$ outside the shaded area of Figure \ref{fig.map}.
The boundary (dashed line) consists of three regions:
\begin{enumerate}
\item $N_{\rm cl} = N_{\rm cl}|_{m_{\rm max} = 20 m_{\sun}} = {\rm const.}$, where $N_{\rm cl}|_{m_{\rm max} = 20 m_{\sun}}$ is the number of stars in the cluster for $m_{\rm max} = 20 m_{\sun}$,
\item $t_{\rm SF} = t_{\star, {\rm max}}$, where $t_{\star, {\rm max}}$ is the lifetime of the most massive star in the cluster,
\item $t_{\rm SF} = t_{\star}|_{m_{\star} = 60 m_{\sun}} = {\rm const.}$, where $t_{\star}|_{m_{\star} = 60 m_{\sun}}$ is the lifetime of stars whose mass is $60 m_{\sun}$.
\end{enumerate}

As $t_{\star}|_{m_{\star} = 60 m_{\sun}} \simeq 4~{\rm Myr}$ (see Figure \ref{fig.LT}), $P_{\rm SN}$ sharply decreases when $t_{\rm SF} \to 4~{\rm Myr}$.
For example, if we set $N_{\rm cl} = 10^{4}$, we obtained that $P_{\rm SN} = 2 \times 10^{-3}$ for $t_{\rm SF} = 4~{\rm Myr}$, $P_{\rm SN} = 0.3$ for $t_{\rm SF} = 5~{\rm Myr}$, and $P_{\rm SN} = 0.7$ for $t_{\rm SF} = 6~{\rm Myr}$.

The condition for $P_{\rm SN} = 0.1$ is not significantly different from the condition for $P_{\rm SN} = 0$ in $N_{\rm cl}$--$t_{\rm SF}$ plane.
This corresponds to the sharp increase of $P_{\rm SN}$ around $P_{\rm SN} = 0$ as shown in Figure \ref{fig.P}, and this is the feature for the model with variable $m_{\rm max}$.
We therefore regard the shaded region as the possible parameters of the Sun's birth cluster.

When we focus on the condition for $P_{\rm SN} > 0.5$ (magenta solid line), the plausible number of stars in the Sun's birth cluster would be larger than $2 \times 10^{3}$ when $t_{\rm SF} < 10~{\rm Myr}$.
Moreover, the plausible number would be $N_{\rm cl} > 2 \times 10^{4}$ when much shorter timescale of $t_{\rm SF} < 5~{\rm Myr}$ is assumed.

\section{Discussion}
\label{sec.discussion}

\subsection{On the duration of star formation in the birth cluster}
\label{sec.duration}

Understanding the duration of star formation in the birth cluster is crucial for determining the plausible $N_{\rm cl}$.
\citet{2021PASJ...73.1074F} studied the formation of massive star cluster that satisfies $N_{\rm cl} > 10^{4}$, and they found that the duration of star formation in star clusters has a significant variation.
For a large cluster formed in a numerical simulation of \citet{2021PASJ...73.1074F}, the duration of star formation is $t_{\rm SF} > 5~{\rm Myr}$, while another small cluster showed a very short duration of approximately 1 million years (see their Figure 25).

The maximum value of $t_{\rm SF}$ would be evaluated from the sum of free-fall timescale and disruption timescale of star-forming giant molecular clouds due to expansion of \ion{H}{II} bubbles \citep[e.g.,][]{2020MNRAS.497.3830F}.
As the efficiency of photoionization feedback strongly depends on the initial surface density of star-forming giant molecular clouds \citep[e.g.,][]{2021MNRAS.506.5512F}, $t_{\rm SF}$ may also depend on the initial surface density, and thus it should depend on the cluster radius.
Future studies on this point is needed to evaluate the physical parameters of the Sun's birth cluster.

%%Krumholtz+ (2019) ARA&A などをみても面密度 Sigma を用いて議論しているので業界では一般的なのかもしれません。

\subsection{Dynamical constraints on the birth cluster}

Not only the existence of SLRs but also dynamical properties of solar system bodies give crucial constraints on $N_{\rm cl}$.
\citet{2009ApJ...696L..13P} and \citet{2010ARA&A..48...47A} reviewed the constraints on the number of stars ($N_{\rm cl}$) and radius ($R_{\rm cl}$) of the Sun's birth cluster.
Close encounters of stars within a distance of 100 au can easily destroy the planetary system, and the sharp truncation of the Kuiper belt at approximately 50 au could be reproduced when the distance of the closest encounter, $b$, is in the range of 150--200 au \citep[e.g.,][]{2001Icar..153..416K}.
Considering the dynamical evaporation of the star cluster, \citet{2001Icar..153..416K} estimated the relation among $b$, $N_{\rm cl}$ and $R_{\rm cl}$ as follows \citep[see also][]{2001Icar..150..151A}:
\begin{equation}
b \sim 200 {\left( \frac{N_{\rm cl}}{2 \times 10^{3}} \right)} {\left( \frac{R_{\rm cl}}{2~{\rm pc}} \right)}^{-1}~{\rm au}.
\end{equation}

\citet{2019MNRAS.482..732C} listed $N_{\rm cl}$ and $R_{\rm cl}$ for various star clusters and star-forming regions which host class II young stellar objects.
They reported that $N_{\rm cl} \simeq 2000$ and $R_{\rm cl} \simeq 1~{\rm pc}$ for the Orion Trapezium cluster, and $N_{\rm cl} \simeq 12700$ and $R_{\rm cl} \simeq 52~{\rm pc}$ for the Lupus cloud star-forming region.
Using these values, we obtained $b$ for these two regions: $b \sim 100~{\rm au}$ for the Orion Trapezium cluster and $b \sim 800~{\rm au}$ for the Lupus cloud star-forming region.
A large cluster with $N_{\rm cl} \sim 10^{4}$ and $R_{\rm cl} \simeq 10~{\rm pc}$ might be plausible from the points of view of dynamical and cosmochemical constraints.

We note that the estimation above is based on the assumption that stars are distributed homogeneously within a cluster.
However, in reality, stars in clusters would be initially distributed in a fractal \citep[e.g.,][]{2004A&A...413..929G}.
\citet{2019A&A...622A..69P} argued that $b$ tends to be small when a cluster with an initial fractal distribution is considered.

\subsection{On the timing of injection event}

There are two major scenarios for the direct injection of SLRs from a nearby CCSN: injection to the molecular cloud core \citep[e.g.,][]{1977Icar...30..447C, 2002ApJ...575.1144V, 2008ApJ...686L.119B, 2012ApJ...745...22G} and to the circumstellar disk \citep[e.g.,][]{1977Icar...32..255C, 2000ApJ...538L.151C, 2007ApJ...662.1268O, 2010ApJ...711..597O}.
Here we briefly review these scenarios.

Molecular cloud cores are the progenitors of protostars, and the gravitational collapse of molecular cloud cores forms stars and their circumstellar disks.
The duration of this stage is typically several $10^{5}$ years, and it corresponds to the class 0/I young stellar objects \citep[e.g.,][]{2011AREPS..39..351D}.
The injection to the molecular cloud core is regarded not only as the event for SLR enrichment but also as the trigger for the collapse of the protosolar molecular cloud core \citep[e.g.,][]{2014MNRAS.444.2884L, 2021ApJ...921..150K}.

Another scenario is the direct injection of dust particles to the earliest phase of the circumstellar disk around the young Sun corresponding to class II young stellar objects.
\citet{2010ApJ...711..597O} found that the injection efficiency depends on the size of dust particles, and 0.1--1 micron sized dust particles could be implanted to the circumstellar disk.
Nearby CCSNe would have a great impact not only on the abundance of SLRs but also the structural evolution of the circumstellar disk.
\citet{2018A&A...616A..85P} numerically investigated interactions between a supernova (both irradiation and blast wave) and the solar circumstellar disk.
They concluded that the observed misalignment between the Sun's equator and the ecliptic plan \citep[e.g.,][]{2005ApJ...621L.153B} and the sharp truncation of the Kuiper belt \citep[e.g.,][]{2001ApJ...549L.241A} could be explained by a nearby CCSN event.

We note that the timing of injection event could also affect the isotopic heterogeneity found in various stable isotopes.
\citet{2021ApJ...908...64F} mentioned that injection of clumpy supernova ejecta into the circumstellar disk around the young Sun could be reasonable from the point of view of the nucleosynthetic chromium isotopic variation recorded in carbonaceous chondrites, although we cannot rule out the possibility that the injection event has occurred in the molecular cloud core stage.

\subsection{Injection of SLRs from Wolf-Rayet winds}

Wolf-Rayet (WR) stars are massive evolved stars whose surface composition has been altered by mass loss and internal mixing.
The minimum initial mass for a star to become a WR star is approximately $25 M_{\sun}$ at the solar metallicity, although the threshold mass depends on its rotation \citep[e.g.,][]{2012A&A...542A..29G, 2014ARA&A..52..487S}.
For nonrotating WR stars, the mass loss due to WR winds starts after their main-sequence and the lifetime in the WR phase is less than 0.5 million years \citep[e.g.,][]{2003A&A...404..975M, 2012A&A...542A..29G}.

WR winds might play an important role as injectors of SLRs \citep[e.g.,][]{1997A&A...321..452A, 2006A&A...453..653A, 2012A&A...545A...4G}.
As WR winds are enriched in $^{26}{\rm Al}$ but depleted in $^{60}{\rm Fe}$, several studies \citep[e.g.,][]{2015ApJ...802...22T} regarded WR stars as one of the most plausible sources for the origin of SLRs in the early solar system (see Appendix \ref{appB}).

There are two types of scenarios for introducing SLRs from WR stars: direct injection to solar nebula by WR winds \citep[e.g.,][]{2019A&A...622A..69P} and enrichment of SLRs before the formation of the parental molecular cloud core \citep[e.g.,][]{2012A&A...545A...4G, 2017ApJ...851..147D}.
In the latter scenario, the Sun would be born in the dense shell around a WR star together with a huge numbers of solar siblings.

We note that the time interval of the start of the WR phase and the following CCSN explosion would be approximately 0.5 million years or less.
This timescale is comparable to the timescale for collapse of the molecular cloud core.
Therefore, we can imagine that both WR wind and CCSN could contribute to the enrichment of SLRs in the early solar system.
We should investigate what would be happen if both WR wind and CCSN occurred within the Sun's birth cluster in future studies.

\subsection{Alternative scenarios without direct injection events}

In this study, we focus on the situation that SLRs in the early solar system are directly injected from nearby sources.
We note, however, that several studies proposed that sequential star formation events and/or self-enrichment in parental molecular clouds may also explain the abundance of SLRs in the early solar system \citep[e.g.,][]{2012A&A...545A...4G, 2015A&A...582A..26G, 2018MNRAS.480.4025F, 2021NatAs...5.1009F, 2022arXiv220311169D}.

Although these scenarios can naturally explain the abundance of SLRs recorded in normal CAIs, we point out that these scenarios would have difficulty in reproducing the coexistence of $^{26}{\rm Al}$-poor unusual CAIs and $^{26}{\rm Al}$-rich normal CAIs.
\citet{2016ApJ...826...22K} performed giant molecular cloud scale numerical simulations that can trace the star formation history and abundance of $^{26}{\rm Al}$ in star-forming gas.
They reported that $^{26}{\rm Al} / ^{27}{\rm Al}$ ratio of accreting gas onto a newborn star is spatially and temporally homogeneous during its class 0/I phase.
This might indicate that all CAIs have same initial value of $^{26}{\rm Al} / ^{27}{\rm Al}$ in the framework of these scenarios; it seems inconsistent with meteoritical evidence.

\citet{2016ApJ...826...22K} and \citet{2020E&PSL.53516088L} speculated that FUN CAIs and PLACs might be formed via selective thermal processing of presolar dust grains.
Their scenario requires that CAIs formed via partial evaporation and recondensation with spatial partitioning; however, whether the CAI formation region satisfies these requirements or not is unclear.
Future studies on the particle dynamics in the innermost region of the solar nebula would be needed.

\section{Summary}

Several pieces of evidence indicate that our solar system has formed within a star cluster.
Injection of SLR-rich dust grains into the solar system from a nearby CCSN is a classical scenario for the origin of SLRs found in meteorites including $^{26}{\rm Al}$ and $^{60}{\rm Fe}$.
Although the majority of primitive CAIs are uniformly enriched in $^{26}{\rm Al}$, some unusual CAIs (i.e., FUN CAIs and PLACs) show the order-of-magnitude lower initial abundance of $^{26}{\rm Al}$.
The coexistence of $^{26}{\rm Al}$-rich and $^{26}{\rm Al}$-poor CAIs is usually interpreted as the evidence of the direct injection of $^{26}{\rm Al}$-enriched dust grains into the solar system in the epoch of CAI formation.
As the duration of CAI formation is approximately $10^{5}$ years, the direct injection event would have occurred in the first $10^{5}$ years of the solar system formation.
Therefore, at least one CCSN would occur in the Sun's birth cluster during its star formation period that would be typically several million years (Figure \ref{fig.1}).

The condition for acquiring at least one CCSN in the birth cluster has been studied in previous studies \citep[e.g.,][]{2010ARA&A..48...47A}.
They evaluated the minimum number of stars in the birth cluster based on the probability for existing at least one massive star whose mass is large enough to trigger a CCSN with ejection of SLR-rich dust grains.
However, those studies did not consider the timing of CCSNe in the cluster, and the existence of massive stars in the cluster does not support the occurrence of CCSN events during its star formation period.

In this paper, we revisited the number of stars in the Sun's birth cluster from the point of view of direct injection of SLRs from a CCSN to the early solar system.
We calculated the probability for acquiring at least one CCSN within the finite duration of star formation in the birth cluster (Figure \ref{fig.P}).
We found that the estimated number of stars in the birth cluster, $N_{\rm cl}$, is significantly larger than that previously considered, especially for the cases that the duration of star formation, $t_{\rm SF}$, is far less than 10 million years.

The plausible number of stars in the Sun's birth cluster would be $N_{\rm cl} > 2 \times 10^{3}$ when $t_{\rm SF} < 12~{\rm Myr}$ (Figure \ref{fig.map}), although it depends on many assumptions.
Moreover, the plausible number would be $N_{\rm cl} > 2 \times 10^{4}$ when much shorter timescale of $t_{\rm SF} < 5~{\rm Myr}$ is assumed.
In contrast, the minimum number of stars without consideration of the timing of explosion, i.e., the lower limit of $N_{\rm cl}$ obtained from a classical way, is approximately 500 (Figure \ref{fig.P_c}).
In other words, our novel constraint is an order of magnitude higher than previous estimates.

Understanding the evolution of the solar nebula in the birth cluster is essential to unveiling how the Earth formed.
The effects of nearby massive stars on disk evolution (e.g., external photoevaporation and stellar feedback from winds and supernovae; \cite{2019MNRAS.482..732C}) should depend on $N_{\rm cl}$, $t_{\rm SF}$, and the geometry of the cluster.
Future studies on the disk evolution around solar-type stars in a large star cluster taking the impact of finite $t_{\rm SF}$ into account would be crucial.

We acknowledge that whether the source of SLRs is a CCSN or a WR star is still under debate.
Moreover, whether a direct injection event contribute to the SLR enrichment of the solar system or not is also controversial.
We discussed these topics in Section \ref{sec.discussion}.
These uncertainty on the origin of SLRs must have a great impact on the evaluation of $N_{\rm cl}$ and $t_{\rm SF}$ for the Sun's birth cluster.
We need to discuss this point in depth in future studies.

\begin{acknowledgements}
%%The anonymous reviewer provided a constructive review that improved this paper.
The authors wish to express their cordial thanks to the referee Anthony G.\ A.\ Brown for constructive comments.
The authors thank Michiko S.\ Fujii, Takashi J.\ Moriya, and Ryota Fukai for useful comments and discussions.
S.A.\ was supported by JSPS KAKENHI Grant No.\ JP20J00598.
E.K.\ was supported by JSPS KAKENHI Grant No.\ 18H05438.
This work was supported by the Publications Committee of NAOJ.
\end{acknowledgements}

\bibliographystyle{aa}
\bibliography{sample}

%%\clearpage
\appendix

\section{Derivation of the relation between $Z_{\rm SN}$ and $P_{\rm SN}$}
\label{appA}

Here we briefly derive the relation between $Z_{\rm SN}$ and $P_{\rm SN}$.
We set the time interval ${\Delta t}$ as ${\Delta t} = t_{\rm SF} / N$, where $N$ is a large integer.
When ${\Delta t}$ is sufficiently small, the probability that no CCSN occur between the time $t$ and $t + {\Delta t}$ is equal to $1 - {p_{\rm SN} {( t )}} {\Delta t}$.
Then the probability for acquiring no CCSN within $t_{\rm SF}$, $1 - P_{\rm SN}$, is given by the following equation:
\begin{equation}
1 - P_{\rm SN} = \lim_{{\Delta t} \to 0} \prod_{n = 0}^{N-1}~{\left[ 1 - {p_{\rm SN} {(n {\Delta t})}} {\Delta t} \right]}.
\end{equation}
By taking the logarithm of both sides, the equation above is converted as follows:
\begin{eqnarray}
\log{\left( 1 - P_{\rm SN} \right)} & = & \lim_{{\Delta t} \to 0} \sum_{n = 0}^{N-1}~\log{\left[ 1 - {p_{\rm SN} {(n {\Delta t})}} {\Delta t} \right]} \nonumber \\ 
                                    & = & - \lim_{{\Delta t} \to 0} \sum_{n = 0}^{N-1}~{p_{\rm SN} {(n {\Delta t})}} {\Delta t} \nonumber \\
                                    & = & - Z_{\rm SN},
\end{eqnarray}
Therefore, we obtained the relation between $Z_{\rm SN}$ and $P_{\rm SN}$:
\begin{equation}
P_{\rm SN} = 1 - \exp{\left( - Z_{\rm SN} \right)}.
\end{equation}

\section{Abundance of $^{60}{\rm Fe}$ in the early solar system}
\label{appB}

The initial abundance of $^{60}{\rm Fe}$ is the key to unveiling the source of SLRs in the early solar system.
It is known that $^{60}{\rm Fe}$ is barely produced by energetic-particle irradiation around the young Sun \citep[e.g.,][]{1998ApJ...506..898L}, and its presence requires an extrasolar contribution from massive stars.

The initial ratio of $^{60}{\rm Fe} / ^{56}{\rm Fe}$ has been investigate in a large number of studies \citep[e.g.,][]{2003ApJ...588L..41T, 2012E&PSL.359..248T, 2015ApJ...802...22T, 2018GeCoA.221..342T, 2018ApJ...857L..15T}.
\citet{2003ApJ...588L..41T} measured troilite grains in least metamorphosed ordinary chondrites and reported that the initial ratio of $^{60}{\rm Fe} / ^{56}{\rm Fe}$ for the solar system would be in the range between $2.8 \times 10^{-7}$ and $4 \times 10^{-7}$.
As the steady-state ratio for the interstellar medium is $2.6 \times 10^{-8}$ \citep{1996ApJ...466L.109W}, they claimed that enrichment of $^{60}{\rm Fe}$-rich materials is required shortly before or during solar system formation.
\citet{2008ApJ...688.1382T} found that the abundance ratio of $^{60}{\rm Fe}$ and $^{26}{\rm Al}$ would be consistent with a injection of ejecta from a CCSN with mixing and fallback \citep[e.g.,][]{2006NuPhA.777..424N}.
The range of the initial mass of the progenitor would be $20$--$40 m_{\sun}$ based on the original analysis of \citet{2008ApJ...688.1382T}, and \citet{2009GeCoA..73.4922H} noted that mixing-fallback process could potentially explain the abundance ratio of SLRs for stars up to $60 m_{\sun}$.

The initial value of $^{60}{\rm Fe} / ^{56}{\rm Fe}$ is still under debate, however.
\citet{2015ApJ...802...22T} measured chondrules in the least metamorphosed ordinary chondrite, Semarkona, and they reported a much lower level of $^{60}{\rm Fe} / ^{56}{\rm Fe}$ \citep[see also][]{2022ApJ...929..107K, 2022ApJ...940...95K}.
The initial abundance obtained from their measurements corresponds to $^{60}{\rm Fe} / ^{56}{\rm Fe}$ ratio of ${(1.01 \pm 0.27)} \times 10^{-8}$, and no injection event is needed to explain this low abundance of $^{60}{\rm Fe}$.
\citet{2018GeCoA.221..342T} also measured chondrules in least metamorphosed ordinary chondrites, and they concluded that the initial $^{60}{\rm Fe} / ^{56}{\rm Fe}$ ratio would be $8.5 \times 10^{-8}$ to $5.1 \times 10^{-7}$, higher than the interstellar medium background.
This is consistent with a scenario that the early solar system was enriched in materials from CCSN ejecta.

\end{document}